\documentclass[12pt,a4paper]{article}
\usepackage{graphicx}
\usepackage{times}
\title{\bf Multiplicity in 5 $M_\odot$ Stars }
\author{Nancy Remage Evans$^1$\\
\vspace{1cm}\\
\normalsize $^1$ Smithsonian Astrophysical Observatory\\ 
}
\date{\mbox{}}
\begin{document}
\maketitle
\pagestyle{empty}
%
%
\def\bull{\vrule height .9ex width .8ex depth -.1ex}
\makeatletter
\def\ps@plain{\let\@mkboth\gobbletwo
\def\@oddhead{}\def\@oddfoot{\hfil\tiny\bull\quad
``The multi-wavelength view of hot, massive stars''; 39$^{\rm th}$ Li\`ege Int.\ Astroph.\ Coll., 12-16 July 2010 \quad\bull}%
\def\@evenhead{}\let\@evenfoot\@oddfoot}
\makeatother
%
%
\def\beginrefer{\section*{References}%
\begin{quotation}\mbox{}\par}
\def\refer#1\par{{\setlength{\parindent}{-\leftmargin}\indent#1\par}}
\def\endrefer{\end{quotation}}
%
%
{\noindent\small{\bf Abstract:} 
Multiwavelength opportunities have provided important
new insights into the properties of binary/multiple  5 $M_\odot$ 
stars.  The combination of cool evolved primaries and 
hot secondaries in Cepheids (geriatric B stars) has yielded 
detailed information about the distribution of mass ratios.
It has also provided a surprisingly high fraction of 
triple systems.  Ground-based radial velocity orbits combined with  
satellite data from Hubble, FUSE, IUE, and Chandra can
provide full information about the systems, including the
masses. In particular, X-ray observations can identify low 
mass companions which are young enough to be physical 
companions.  These multiwavelength observations provide important 
tests for star formation scenarios including diffenences
between high and low mass results and differences 
between close and wide binaries.

}
%
%
\section{Introduction}

The multiplicity of stars provides important clues to star formation
processes, and in some cases is the most important determinant of
their future.  For 5 $M_\odot$ stars, multiwavelength observations have
provided detailed information about their multiplicity and also 
their binary/multiple properties.  Specifically Cepheids (post-main
sequence He burning stars in the ``blue loop"  phase of evolutionary
tracks) can be used  to provide such information in novel ways.

\subsection{Multiplicity}

Because these cool supergiants often have hot main sequence
companions, ultraviolet spectra (HST and IUE) provide a spectrum of
the companion uncontaminated by the light of the primary (Evans 1995).
 This has been used to determine Cepheid masses (Evans, Carpenter, 
Robinson, et al. 2005,
 and references therein).  It has also been a particularly valuable
way to identify systems which are  not simply binaries, but triple
systems  (Evans, et al. 2005).  In fact, without some direct
information about the secondary, one cannot be confident in general that a system
contains only two stars.  

Triple systems among Cepheids have been identified in many ways (see
Evans, et al. 2005).  W Sgr provides one example.  It was known both
to be a binary system  with an orbit (Babel, Burki, Mayor, 
et al. 1989) and to have a hot companion
(B\"ohm-Vitense and Proffitt 1985; Evans 1991).  However, the 
combination was not consistent with a reasonable
mass for the Cepheid (see Evans, Massa, and Proffitt 
2009).  The solution came from
an HST STIS (Space Telescope Imaging Spectrograph) spectrum (Fig 1).  The
hottest star in the system is resolved from the Cepheid and its
spectroscopic binary companion.  This new insight results in a
reasonable upper limit to the Cepheid mass from the astrometric
orbit of the Cepheid (Benedict, McArthur, Feast, et al. 2007).

\begin{figure}[h]
\centering
\includegraphics[width=8cm]{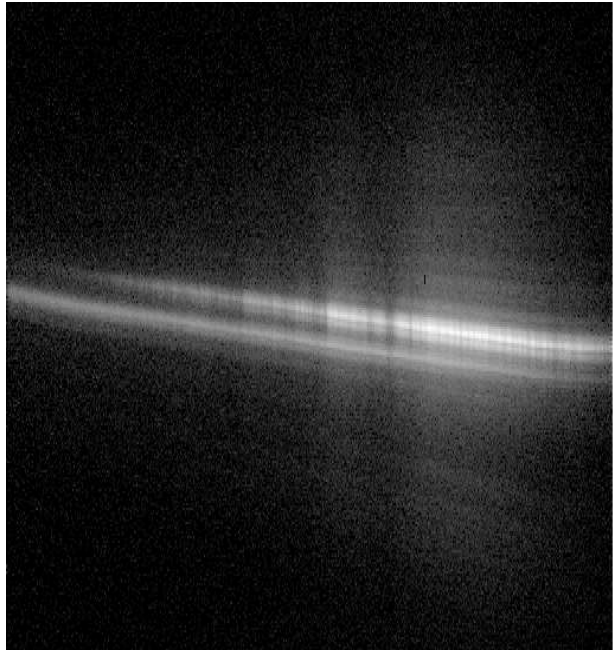}
\caption{The flat-fielded HST STIS image of W Sgr.  Wavelength
  increases to the right from about 1796 to 3382 \AA.  The cooler
  component (Cepheid + spectroscopic binary companion) 
is the upper one in the image and clearly separated from
  the hotter component below which extends much further toward shorter
  wavelengths. The spectrum flux is on a log scale.  The strongest feature
  in the Cepheid spectrum is Mg II 2800\AA. 
Reprinted from Evans, et al. 2009, AJ, 137,3700.\label{fig_1}}
\end{figure}

Another example of the  detection of a third star from a spectrum of the
secondary is provided by  SU Cyg.  The hot companion of the Cepheid is
known to be a binary from ultraviolet velocities.  However, in
addition, it has a strong GaII feature at 1414 \AA, indicating that it
is a HgMn star (Wahlgren and Evans 1998).  Since these chemically
peculiar stars require very slow rotation to facilitate  element
diffusion, they are found in short period binaries, whose orbit and rotation
have been tidally locked.  The GaII feature is easily identifiable on a
low resolution spectrum, indicating that the companion is itself a binary itself.

In order to determine the fraction of well-studied binary Cepheid
systems which are in fact triple we (Evans et al 2005) have
compiled a list of 18 Cepheids with orbits which have an ultraviolet
spectrum of the companion.  Of these, 44\% (possibly 50\%) are
triples, a very high fraction for 5 $M_\odot$ stars.

\subsection{Mass Ratio Distribution}

Ultraviolet spectra also provide very precise spectral types of the
companion, from which masses can be inferred.  The mass ratios from
the companion mass and a mass inferred for the Cepheids are shown in Fig. 2.
(See Evans [1995] for a full discussion of Cepheid mass ratios
and completeness.)  Note that the Cepheid sample includes only systems
with orbital periods longer than a year, since shorter period systems
would have undergone Roche lobe overflow before the primary became a
supergiant.  For comparison, the mass ratio distribution of solar
mass stars from Duquennoy and Mayor (1991, hereafter referred to as DM) is shown.  The
distributions are very similar, despite the fact that the Cepheid primaries are 5
times as massive as the DM primaries.  For comparison, the mass ratio
distribution of O stars from recent studies is shown (Rauw, Naz\'e, 
Fern\'andes Laj\'us,  et
al. 2009; Sana, Gosset, Naz\'e,  et al. 2008; Sana, Gosset, 
and Evans 2009; Kiminki, Kobulnicky, Gilbert, et al. 2009).
 It contains a population of equal mass binaries and falls
off for low mass companions.  This poses the question of whether
the mass ratio (q) distribution is a function of separation and/or
mass, and also the role of incompleteness.  That is, one explanation
for the difference in Fig. 2 between Cepheids and O stars may be 
that short period binaries in the  O star sample are
more likely to have equal mass binaries than the remaining longer
period (wider separation) binaries in
the Cepheid sample.

\begin{figure}[h]
\centering
\includegraphics[width=8cm]{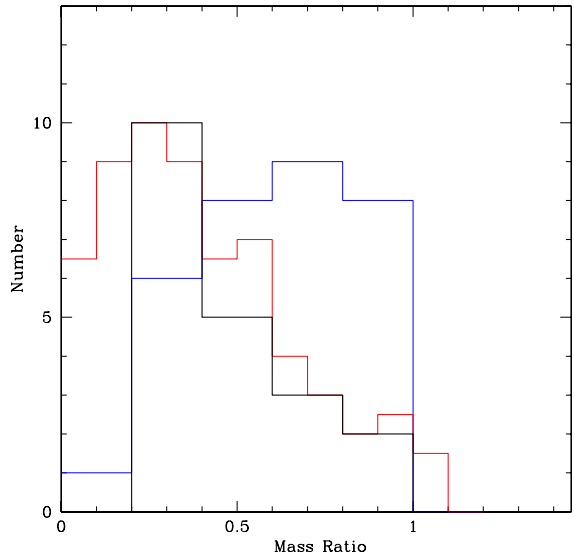}
\caption{The distribution of mass ratios q = M$_2$/ M$_1$.  The black
  line is the Cepheid sample (Evans 1995); the red line is the solar mass sample
  from DM; the blue line is the O star sample.
\label{fig_1}}
\end{figure}

\subsection{Low Mass Companions}

One important aspect of binary properties is the prevalence of low
mass companions for 5 $M_\odot$ primaries.  These are, of course, the
most difficult to detect both by photometric and spectroscopic
(radial velocity) techniques.  We are exploring the use of X-rays to
improve this situation.  Comparable mass main sequence stars (late B
stars) do not in general produce X-rays.  Low mass stars
(spectral types mid F through K) young
enough to be Cepheid/late B star companions (typically 50 Myr old)
produce copious X-rays.  M stars
are weaker X-ray sources; older field stars are also much weaker 
X-ray sources.  A list was drawn up of  B3 to A0 stars in Tr 16 with  proper
motions indicating cluster membership (Cudworth, Martin, and
DeGioia-Eastwood 1993).
Fig. 3 shows the sample with X-ray detections from a Chandra image 
 (Evans, DeGioia-Eastwood, Gagn\'e, et al. 2011; Townsley, Broos, Corcoran,
et al. 2011; Albacete-Colombo, Damiani, Micela, et al 2008). 
(Lines show the ZAMS for 2.3 kpc with E(B-V) = 0.45, 0.55, and 0.65 mag.)
39\% of the late B stars are detected as X-ray sources, indicating that they
have low mass companions.  See Evans, et
al. (2011) for a complete discussion, including the X-ray detection
fraction.

\begin{figure}[h]
\centering
\includegraphics[width=8cm]{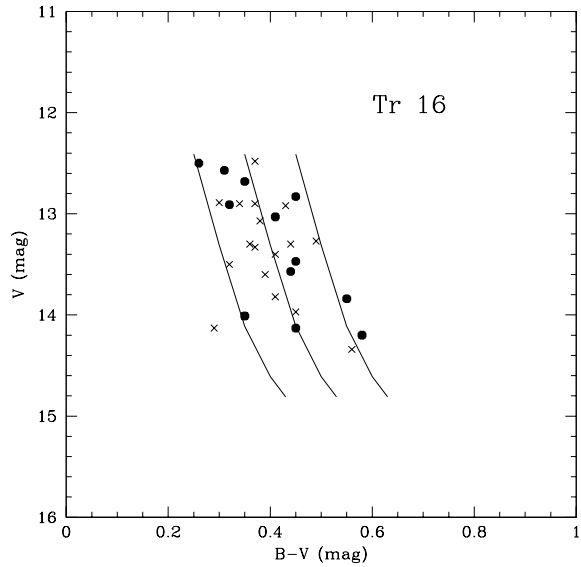}
\caption{Late B stars in Tr 16 (from Evans, et al. 2011).  Dots are detected in X-rays; x's are
  not detected.  The lines are the ZAMS for 2.3 kpc and an appropriate
  range of B-V corresponding to E(B-V) = 0.45, 0.55, and 0.65 mag.
\label{fig_1}}
\end{figure}

\subsection{Discussion}

As shown in the previous sections, multiwavelength observations have
provided considerable new insight into the binary/multiple properties
of 5 $M_\odot$ stars.  Here we discuss some remaining issues about multiplicity.

Direct observations of companions, provided particularly in
ultraviolet studies by HST and IUE, find a high fraction of triple
systems (at least 44\%).  Even this, however, may not be the total
count of system members. For instance, in some systems with low mass 
companions, the low mass companion might itself be a binary.
 
The distribution of mass ratios (Fig. 2) strongly favors unequal mass
companions. This  pertains, however, only to Cepheids with separations 
(A sin i) larger than about 0.4 AU and periods longer than a year
(Sugars and Evans 1996).  This is at least part of the reason for the
difference between Cepheid mass ratios and those of O stars in
Fig. 2.  

A way to identify low mass companions of late B stars (comparable in mass
to Cepheids) using X-ray images is shown in Fig. 3. How complete is
our knowledge of the total binary/multiple fraction for 5 $M_\odot$ stars? 
The fraction Cepheids with companions hotter than mid A spectral type
is known from an IUE survey of the 76 Cepheids brighter than 8$^{th}$
mag (Evans 1992). 21\% were found to have companions (which rises to 34\% after a
statistical correction was included  based on stars with known orbital
motion). The ``Chandra fraction'' of low mass companions and the ``IUE
fraction'' should have relatively little overlap, and hence are
approximately additive.  Furthermore, the results tentatively suggest
that the steep rise in secondary mass frequency seen in the DM solar
mass stars from 1 to 0.2 $M_\odot$ stars is mimiced in the 5$M_\odot$ 
mass ratio distribution. However, a drop off at the lowest mass
ratios for the O stars (if confirmed) would imply a different
frequency of 1 $M_\odot$ companions for O stars and solar mass stars. 
 That is, the {\it mass ratio} not the {\it mass} may be the
important parameter at the time in star formation scenarios  when the
q distribution is determined.

For comparison, the recent study by Mason, Hartkopf, Gies, 
et al. (2009) of combined
spectroscopic binary velocity results with interferometry for O and B
stars. (X-ray identification of young companions cannot be used, 
since O and early B stars in this sample  
are intrinsic X-ray producers.)  They find a
binary fraction of 66\% for O stars.  Their lists, of course, are not
likely to contain low mass companions since small mass ratios and large magnitude
differences make these companions very difficult to detect.


%
%
\section*{Acknowledgements}

Funding for this work was provided by 
Chandra X-ray Center NASA Contract NAS8-39073.      

%
%

\vskip .5truein

\footnotesize
\beginrefer

\refer Albacete-Colombo, J. F., Damiani, F., Micela, G., Sciortino, S., and
Harnden, F. R., Jr. 2008, A\&A, 490, 1055 

\refer Babel, J., Burki, G., Mayor, M., Waelkens, C., and Chmielewski,
Y. 1989, A\&A, 216, 125

\refer Benedict, G. F., McArthur, B. F., Feast, M. W., 
et al. 2007, AJ, 133, 1810

\refer B\"ohm-Vitense, E. and Proffitt, C. 1985, ApJ, 296, 175

\refer Cudworth, K. M., Martin, S. C., and
  DeGioia-Eastwood, K. 1993, AJ, 105, 1822

\refer Duquennoy, A. \&   Mayor, M. 1991, A\&Ap, 248, 485 (DM)

\refer Evans, N. R. 1991, ApJ 372, 597

\refer Evans, N. R. 1992, ApJ, 384, 220

\refer Evans, N. R. 1995, ApJ, 445, 393

\refer Evans, N. R., Carpenter, K. G.,
  Robinson, R., Kienzle, F., and Dekas, A. E. 2005, AJ, 130, 789

\refer Evans, N. R., Massa, D., and Proffitt, C. 2009, AJ, 137,3700

\refer Evans, N. R., DeGioia-Eastwood, K,
  Gagne, M., et al.  2011, ApJS, preprint

\refer Kiminki, D. C., Kobulnicky, H. A., Gilbert, I., Bird, S., and
Chunev, G. 2009, AJ, 137, 4608  

\refer Mason, B. D., Hartkopf, W. I., Gies, D. R., Henry, T. J., and Helsel,
J. W. 2009, AJ, 137, 3358

\refer Rauw, G. Naz\'e, Y., Fern\'andez Laj\'us, E., Lanotte, A. A.,
Solivella, G. R., Sana, H., and Gosset, E. 2009, MNRAS, 398, 1582 

\refer Sana, H., Gosset, E., Naz\'e, Y., Rauw, G., and Linder, N. 2008, MNRAS, 386, 447 

\refer Sana, H., Gosset, E., and Evans, C. J. 2009, MNRAS, 400, 1479  

\refer Sugars, B. J. A. and Evans, N. R. 1996, AJ, 112, 1670

\refer Townsley, L. K., Broos, P. S., Corcoran, M. F.  et al. 2011, ApJS, preprint 

\refer Wahlgren, G. M. and Evans, N. R. 1998, A\&A, 332, L33

\endrefer           
\end{document}